\documentclass[aps,prl,twocolumn,showpacs]{revtex4-2}

\usepackage[english]{babel}
\usepackage[utf8]{inputenc}
\usepackage[colorlinks=true,citecolor=blue]{hyperref}
\usepackage{physics}
\usepackage{amssymb,amsmath,amsbsy,bm,amscd}
\usepackage[shortlabels,inline]{enumitem}
\usepackage{epsf,epsfig}
\usepackage{graphicx}
\usepackage{hyperref} 
\usepackage{url}
\usepackage{natbib}

\hypersetup{
    colorlinks,
    linkcolor={red},
    citecolor={red},
    urlcolor={blue}
}

\begin{document}
\title{Subsystem Statistics and Conditional Self-Similarity of Random Quantum States}
\author{Sangchul Oh}
\email{sangchul.oh@siu.edu}
\affiliation{School of Physics and Applied Physics, Southern Illinois University, 
Carbondale, IL 62901, USA}
\date{\today}

\begin{abstract}
We analytically derive the bit-string probability distributions of subsystems of random pure states 
and depolarized random states using the Dirichlet distribution. We identify the exact Beta 
distribution as the universal statistical law of random quantum states, providing a unified 
finite-size description of full-system, subsystem, and conditional statistics. In the presence of 
depolarizing noise, these distributions are scaled and shifted by the noise strength, producing 
a noise-induced gap in their support. Remarkably, we prove that random states exhibit exact 
conditional self-similarity: the distribution of subsystem bit-string probabilities conditioned 
on specific outcomes of the complementary subsystem is identical to that of the full system. 
This hidden scale invariance enables the exact restoration of the full-system statistics from 
the marginalized Beta distribution via post-selection, and persists under depolarizing noise. 
Our results uncover a fundamental symmetry of Hilbert space and provide a scalable, rigorous 
framework for validating random circuit sampling via subsystem or conditional cross-entropy 
benchmarking.
\end{abstract}
\maketitle

A random quantum state is chosen uniformly at random from a Hilbert space. While simply defined, 
random states have novel properties. They reside near the equator of a unit sphere in a Hilbert 
space, called the concentration of measure~\cite{Ledoux2001,Hayden2004}. A subsystem in a random 
pure state is highly entangled~\cite{Page1993a}. A single random state represents an ensemble of 
random states, that is called typicality~\cite{Popescu2006,Reimann2008}. The distribution of 
bit-string probabilities in a random pure state obeys the exponential distribution~\cite{Kus1988}. 
Random quantum states are known to describe a wide variety of physical systems such as chaotic 
quantum systems~\cite{Haake2010}, an eigenstate of a circular unitary matrix in random matrix 
theory~\cite{Broady1981,Kus1988,Mehta2004}, black holes~\cite{Page1993b,Hayden2007}, and random 
circuit sampling on a quantum computer~\cite{Boixo2018,Arute2019,Bouland2019,Wu2021,Zhu2022,
Hangleiter2023,Morvan2024,Gao2025,Liu2025}.

Random circuit sampling is a task to sample bit-strings from a random state generated by random 
quantum circuits. It is regarded as one of the most promising quantum algorithms demonstrating 
quantum advantage on contemporary noisy quantum processors. Random circuit sampling has been 
implemented with 53, 67, 70 superconducting qubits by the Google team~\cite{Arute2019,Morvan2024}, 
with  56, 60, and 83 superconducting qubits by the USTC collaborators~\cite{Wu2021,Zhu2022,Gao2025}, 
with 56 trapped-ion qubits by Quantinuum~\cite{Liu2025}. However, the verification of random 
circuit sampling is challenging as the number of qubits increases. Reconstructing the full 
exponential distribution from sampled bit-strings is computationally formidable for large systems. 
Furthermore, the distribution of bit-string probabilities of noisy qubits is unknown. While the 
linear cross-entropy benchmarking (XEB)~\cite{Boixo2018,Bouland2019,Aaronson2020} requires less 
samples than full distribution reconstruction, it still relies on calculating ideal probabilities 
of finding bit-strings, which limits its scalability.

In this paper, we derive the exact distribution of bit-string probabilities of subsystems in 
a random pure and a depolarized random state-identifying the Beta distribution as the marginal 
distribution of the Dirichlet distribution. This framework provide a unified description of 
statistics of the full system and its subsystems, even conditional subsystems. Under depolarizing 
noise, the corresponding distributions are shifted and scaled by the strength of depolarizing noise, 
resulting in a characteristic depolarizing gap in their support. We prove that a random state exhibits 
{\it conditional self-similarity}: the distribution of probabilities of bit-strings conditioned on 
a subset of bits is statistically identical to that of a full system. This conditional scale 
invariance uncovers the hidden symmetry in a random quantum state. Leveraging this structure, 
we introduce the subsystem or conditional XEB as a scalable verification method for random circuit 
sampling.

\paragraph{Full system statistics of a random pure state.---} We begin with deriving the well-known 
exponential distribution of bit-string probabilities of a random pure state from the Dirichlet 
distribution. A Haar-random pure state of $n$ qubits is written as
\begin{align}
\ket{\psi} = \sum_{i=0}^{N-1} c_i \ket{i}\,,
\label{Eq:random_state}
\end{align}
where $c_i = z_i/\sqrt{\sum_j |z_j|^2}$ and $z_i\sim {\cal CN}(0,1)$ are independent complex normal 
random variables. Here, $N=2^n$ is the dimension of the Hilbert space of $n$ qubits.  
The bit-string probabilities are written as $p_i \equiv |\braket{i}{\psi}|^2 = |c_i|^2$.
Since $|z_i|^2$ follow the Gamma distribution, $|z_i|^2\sim \text{Gamma}(1,1)$, a bit-string 
probability vector $\boldsymbol{p} = (p_0,p_1,\dots,p_{N-1})$ for a random pure state follows the 
flat Dirichlet distribution with all $N$ parameters $\boldsymbol{\alpha}=(1,\cdots,1)$
\begin{align}
\boldsymbol{p} \;\sim\;\text{Dir}(1,1,\cdots,1)\,,
\label{Eq:Dirichlet}
\end{align}
where $\sum_i p_i =1$ and the support is the $N-1$ dimensional 
simplex~\cite{Balakrishnan2004,Zyczkowski2001,Bengtsson2007}. 
The marginal distribution of any single component $p$ of $\boldsymbol{p}$, obtained by integrating 
over the remaining components, follows the Beta distribution
\begin{align}
p \;&\sim\;\text{Beta}(1,N-1) \,,
\end{align}
with its probability density function (PDF) 
\begin{align}
f_{\rm Beta}(p)&= (N-1)\left( 1 -p \right)^{N-2} \,.
\label{Eq:Full_Beta_p}
\end{align}
The PDF with the scaled variable $x\equiv Np$ becomes
\begin{align}
f_{\rm Beta}(x) = \frac{N-1}{N}\left( 1 -\frac{x}{N} \right)^{N-2} \,,
\label{Eq:Full_Beta_x}
\end{align}
where $0\le x<N$ and $1/N$ accounts for the Jacobian. In the limit of large $N$, it becomes
the exponential distribution~\cite{Kus1988}
\begin{align}
f_{\text{Exp}}(x) = e^{-x} \,.
\label{Eq:exponential}
\end{align}

\begin{figure}[t]
\includegraphics[width=0.45\textwidth]{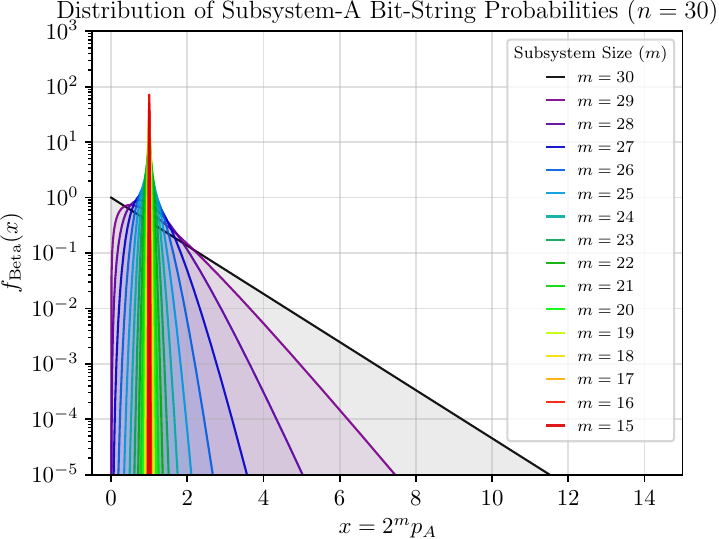}
\caption{\label{Fig1} Plot of Eq.~(\ref{Eq:Sub_Beta_x}), the Beta distributions of subsystem 
bit-string probabilities in a random pure state as a function of $x = Mp_A$ 
for various subsystem sizes.  Here the total number of qubits is $n=30$.}
\end{figure}

\paragraph{Subsystem statistics of a random pure state.---} 
We generalize the Beta distribution of the full-system above to that of arbitrary subsystems.
Consider partitioning the full system of $n$ qubits into two subsystems $A$ with $m$ qubits and $B$ 
with $k$ qubits so $n =m + k$. The corresponding Hilbert space dimension factorizes as $N=M\cdot K$, 
where $M= 2^{m}$ for the dimension of subsystem $A$ and $K=2^{k}$ for the dimension subsystem $B$. 
The bit-string $j$ of the full system is represented by a pair $(y,z)$ where $y = 0, \dots,M-1$ 
and $z = 0,\dots,K-1$, so the probability $p_j$ is denoted by $p(y,z)$. The probability of finding 
a bit-string $y$ of a subsystem $A$ is written as the marginal distribution of $p(y,z)$
\begin{align}
p_A(y) = \bra{y}\rho_A\ket{y}
       = \sum_{z}p(y,z)\,,
\end{align}
where $\rho_A = \tr_B(\ketbra{\psi}{\psi})$ is the reduced density operator of a subsystem $A$ 
for a random pure state $\ket{\psi}$. Subsystem bit-strings may be obtained either by partially 
measuring $m$ qubits of an $n$-qubit random pure state or by taking only $m$-bit substrings 
from full measurement outcomes. 

The probability vector $\boldsymbol{p}=(p(y,z))$ of the full system follows the Dirichlet 
distribution, Eq.~(\ref{Eq:Dirichlet}). The aggregation property of the Dirichlet distribution
states that summing specific components of a Dirichlet-distributed vector yields a new vector that 
is also Dirichlet-distributed, with appropriately combined parameters~\cite{Kotz2019}. 
Thus the probability vector 
$\boldsymbol{p}_A\equiv (p_A(0),\dots,p_A(M-1))$ of a subsystem $A$ follows the Dirichlet 
distribution with $M$ parameters $\bm{\alpha} = (K,\cdots,K)$,
\begin{align}
\boldsymbol{p}_A \sim {\rm Dir}(K,\dots,K)\,.
\label{Eq:partial_Dirichlet}
\end{align}
The marginal distribution of any single component $p_A$ of $\boldsymbol{p}_A$ is given by 
the Beta distribution and its PDF
\begin{align}
p_A &\sim {\rm Beta}(K, N-K) \,,\\[11pt]
f_{\rm Beta}(p_A)  &=\frac{ p_A^{K-1} (1-p_A)^{N-K-1}}{B(K,N-K)}  \,.
\label{Eq:Sub_Beta_p}
\end{align}
Here the Beta function is related to the Gamma function, $B(a,b) = \Gamma(a)\Gamma(b)/\Gamma(a+b)$.
The two shape parameters of the Beta distribution are $K-1$ and $N-K-1$, respectively. 
The PDF with the scaled variable $x\equiv Mp_A$ is written as
\begin{align}
f_{\text{Beta}}(x) 
=\frac{\left(\frac{x}{M}\right)^{K-1}\left(1-\frac{x}{M}\right)^{N-K-1}}{M\cdot B(K,N-K)} \,.
\label{Eq:Sub_Beta_x}
\end{align}
The full system distribution, Eq. (\ref{Eq:Sub_Beta_p}), is the special case of 
Eq. (\ref{Eq:Full_Beta_p}) when system $A$ is equal to the full system so $K=1$, i.e., $k=0$. 

In the limit of large $N$, the PDF approaches the Gamma distribution with shape parameter $K$ and 
scale parameter $1/K$
\begin{align}
f_{\rm Gamma}(x) = \frac{K^{K}}{\Gamma(K)}\,x^{K -1}\,e^{-xK} \,.
\label{Eq:Gamma_distribution}
\end{align}
For the full system, i.e., $K=1$, Eq.~(\ref{Eq:Gamma_distribution}) becomes the exponential 
distribution, Eq. (\ref{Eq:exponential}). If $m \ll k$, i.e., a measured system is 
a tiny fraction of a large system in a random pure state, the PDF becomes the sharp peaked Gaussian 
form centered at $x=1$ with variance $\sigma^2 = 1/K$. Fig.~\ref{Fig1} shows how the Beta distribution 
of bit-string probabilities in a random pure state changes as the size of an unmeasured subsystem 
$k$, ranging from the exponential distribution for the full system the delta-like distribution
for tiny systems.

\paragraph{Full system statistics of a depolarized random state.---} 
The distribution of bit string probabilities obtained from random circuit sampling on a noisy 
quantum computer deviates from the ideal Beta distribution for a random pure state.
In the presence of 
a globally depolarizing channel, the density operator is a mixture of a pure random state 
$\ket{\psi}$ and the fully mixed state $\mathbb{I}/N$
\begin{align}
\rho = (1-\lambda)\ketbra{\psi}{\psi} + \lambda\frac{\mathbb{I}}{N} \,,
\label{Eq:depolarized_state}
\end{align}
where $\lambda$ is the depolarizing noise strength and $N=2^n$. Eq.~(\ref{Eq:depolarized_state}) 
implies that depolarizing noise acts as an affine transformation on the Dirichlet distribution. 
Thus, the marginal distribution is given by
\begin{align}
\tilde{p} \sim (1-\lambda)p + \frac{\lambda}{N}\,,
\label{Eq:depolarized_state_affine}
\end{align}
where $p\sim {\rm Beta}(1,N-1)$ for a random pure state. After the affine transformation of 
the Dirichlet distribution, the PDF, Eq.~(\ref{Eq:Full_Beta_p}), is transformed as
\begin{align}
f(p;\lambda) = \frac{N-1}{(1-\lambda)N} \left(1 -\frac{p-\lambda}{(1-\lambda)N}\right)^{N-2}\,,
\label{Eq:scaled_shifted}
\end{align}
where the support of the distribution is $p\in [{\lambda}/{N}, 1]$.
In the limit of large $N$, using the scaled variable $x= Np$, Eq.~(\ref{Eq:scaled_shifted}) 
becomes the scaled-shifted exponential distribution
\begin{align}
f_{\rm Exp}(x;\lambda) 
= \begin{cases}
\frac{1}{1-\lambda}\exp\left(-\frac{x-\lambda}{1-\lambda}\right) &\text{if $x > \lambda$} \\[11pt]
0 &\text{if $0\le x <\lambda$} 
\end{cases}\,.
\label{Eq:scaled_shifted_exponential}
\end{align}

\begin{figure}[t]
\includegraphics[width=0.45\textwidth]{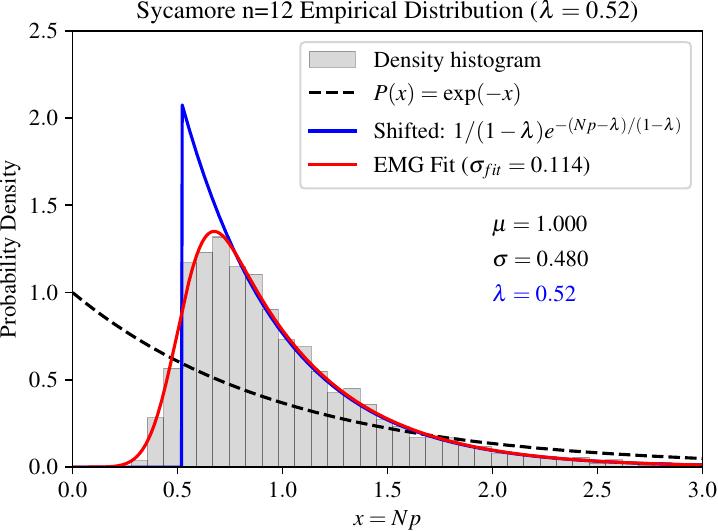}
\caption{\label{Fig:depol} The empirical distribution of bit-string probabilities for Google Sycamore 
with $n=12$ qubit (histogram), the ideal exponential distribution (dotted black line), the scaled-shifted 
exponential distribution, Eq.~(\ref{Eq:scaled_shifted_exponential}) (blue solid line),
and the exponentially modified distribution (red solid line) are plotted. The strengths of depolarizing 
noise is $\lambda= 0.52$.}
\label{Fig2}
\end{figure}

Eq.~(\ref{Eq:scaled_shifted_exponential}) shows that the depolarizing noise shifts the exponential 
distribution, Eq.~(\ref{Eq:exponential}), to the right by the strength of depolarizing noise $\lambda$ 
and enhance the maximum height of the distribution to $1/(1-\lambda)$. Note that the scaled-shifted 
exponential distribution is zero for $0\le x< \lambda$. This region represents a {\it depolarizing gap} 
in the bit-string probability distribution as depicted by the blue solid curve in Fig.~\ref{Fig2}.
Eq.~(\ref{Eq:scaled_shifted_exponential}) supports the numerical simulation of random circuit sampling 
for 20 qubits in Ref.~\cite{Boixo2018}. 

While global depolarization results in the scaled-shifted exponential distribution, 
Eq.~(\ref{Eq:scaled_shifted_exponential}), the empirical distribution for $n=12$ qubits with 
Google Sycamore data~\cite{Martinis2022} exhibits a smooth tail rather than the sharp cutoff predicted 
at the depolarizing gap $x=\lambda$, as shown in Fig.~\ref{Fig2}. This deviation suggests the presence 
of other noise effects beyond the global depolarizing channel. This smoothing may arise from local 
readout errors and coherent control fluctuations. Unlike depolarizing noise, which mixes the state with 
the identity (through an affine transformation), readout errors effectively convolve the ideal 
probability distribution with a noise kernel (typically Gaussian). This convolution smears the sharp 
boundary of the depolarizing gap, pushing probability mass into the theoretically forbidden region 
$0 \le x < \lambda$. Consequently, the empirical distribution may be better described by an 
exponentially modified gauss Ian, where the Gaussian component accounts for these diffusive noise 
processes filling the gap~\cite{Oh2022b,Oh2026}.

\paragraph{Subsystem statistics of a depolarized random state.---} It is straightforward to obtain 
the distribution of subsystem bit-string probabilities under global 
depolarizing noise by combining the two facts of the Dirichlet distribution: 
the subsystem distribution results from the aggregation property of the Dirichlet distribution and
the depolarizing noise result in the affine transformation of the Dirichlet distribution.  
With the scaled variable $x = Mp_A$, the PDF is given by the scaled-shifted Beta distribution
\begin{align}
f_{\rm Beta}(x;\lambda) 
= \frac{1}{(1-\lambda)M}f_{\rm Beta}\left(\frac{x-\lambda}{(1-\lambda)M}\right)\,,
\end{align}
where $x \in [ \lambda, (1-\lambda)M + \lambda]$.

\begin{figure*}[ht]
\includegraphics[width=0.32\textwidth]{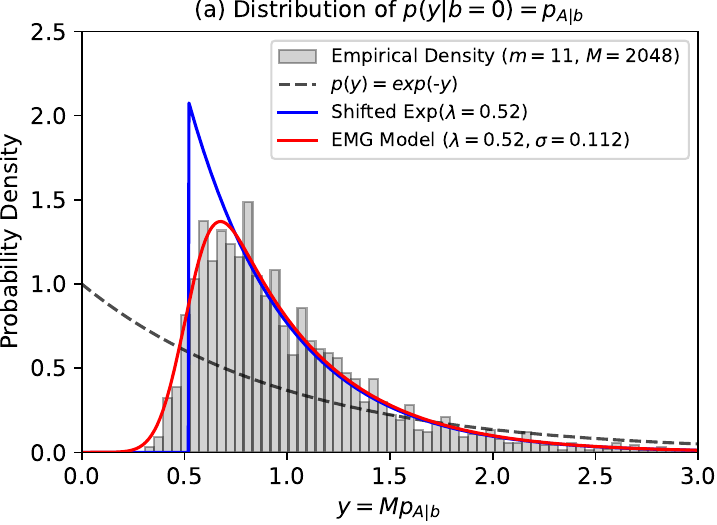}\quad
\includegraphics[width=0.32\textwidth]{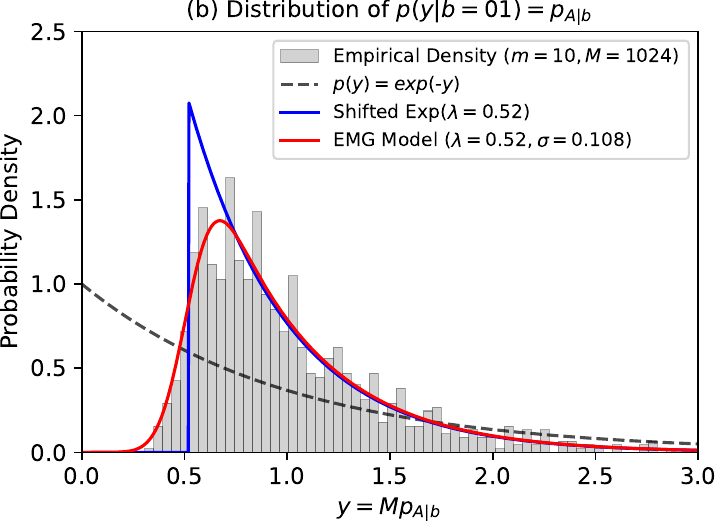}\quad
\includegraphics[width=0.32\textwidth]{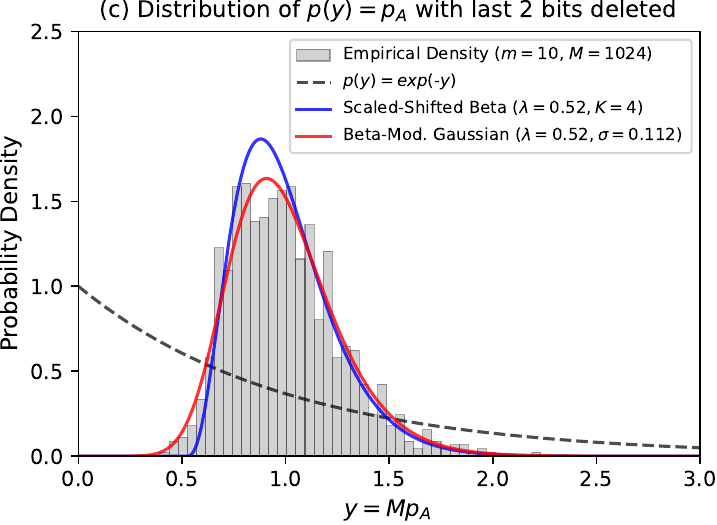}
\caption{\label{Fig:conditional} Using Google Sycamore data for $n=12$ qubits,   
the empirical distributions are plotted for conditional probabilities (a) $p(y|a=0)$, (b) 
$p(y|a=01)$, and (c) the marginal probability $p(y)$ after tracing out last two bits.}
\end{figure*}

\paragraph{Conditional self-similarity of a random state.---} The marginal distribution of 
subsystem $A$, Eq.~{(\ref{Eq:Sub_Beta_p})}, obtained tracing out subsystem $B$ deviates from 
that of the full system, Eq.~{(\ref{Eq:Full_Beta_p})}. We prove that the distribution of bit-string 
probabilities of subsystem $A$ conditioned by bit-strings of system $B$ is statistically identical 
to that of the full system. We call this property {\it conditional scale invariance} or 
{\it conditional self-similarity} that uncovers the hidden fundamental symmetry inherent in the 
Hilbert. This conditional self-similarity is mathematically exact and holds regardless of subsystem 
sizes.

For a fixed $z=b$, the conditional probability of system $A$ is given by
\begin{align}
p(y|b) = \frac{p(y,b)}{\sum_{y'} p(y',b)} = \frac{p(y,b)}{p(b)}\,.
\end{align}
Recalling a random pure state, Eq.~(\ref{Eq:random_state}), 
$\ket{\psi} = \frac{1}{\sqrt{S_G}} \sum_j z_j \ket{j}$, the global normalization factor 
(or global intensity) $S_G$ is the sum of $N$ i.i.d. Gamma random 
variables $X_j \equiv |z_j|^2 \sim {\rm Gamma}(1,1)$
\begin{align}
S_G = \sum_{j} X_j = \sum_{y,z} X_{y,z} \sim {\rm Gamma}(N,1)\,.
\end{align}
The joint probability $p(y,z)=|\braket{y,z}{\psi}|^2$ reads
\begin{align}
p(y,z) = \frac{X_{y,z}}{S_G} \sim {\rm Beta}(1,N-1)\,.
\end{align}
Summing over the subsystem $A$ yield the marginal probability $p(b)$ of the system $B$ 
\begin{align}
p(b) = \sum_y p(y,b) = \sum_y \frac{X_{y,b}}{S_G} = \frac{S_{A|b}}{S_G}\,,
\end{align}
where $S_{A|b} = \sum_{y} X_{y,b}$ is the local normalization factor (the local sum of 
intensities) for the $b$-th branch. Thus, the conditional probability is 
written as
\begin{align}
p(y|b) = \frac{p(y,b)}{p(b)} =\frac{X_{y,b}/S_G}{S_{A|b}/S_G} =  X_{y,b}/S_{A|b}\,,
\end{align}
where the cancel;ation of the global scale $S_G$ uncovers a hidden self-similarity. According to 
Lukacs' Theorem~\cite{Lukacs1955}, for independent Gamma variables, the ratio $X_{y,b}/S_{A|b}$ is statistically 
independent of its sum $S_{A|b}$. This independence, rooted in the neutrality of the Dirichlet 
distribution~\cite{Connor1969,Kotz2019}, ensures that the $M$-component conditional vector 
$\mathbf{p}(y|b) = (p(y_0|b), \dots, p(y_{M-1}|b))$ is itself 
a flat Dirichlet random vector, ${\bm p}(y|b) \sim {\rm Dir}(1^M)$.
Consequently, the marginal distribution of the conditional probabilities $p_{A|b}$ 
exactly restores the universal law:
\begin{align}
p_{A|b} &\sim {\rm Beta}(1,M-1)\,,
\end{align}
and its conditional PDF is  
\begin{align}
f(p_{A|b}) & = (M-1)\left(1-p_{A|b}\right)^{M-2}\,.
\label{Eq:Cond_Beta_p}
\end{align}
This result is mathematically identical to the full-system distribution (Eq.~\ref{Eq:Full_Beta_p}) 
under mapping $N \to M$ and $p\to p_{A|b}$. This conditional scale invariance is exact and 
demonstrate the universal laws of Hilbert space in the sense that a full system statistics of 
a random quantum states is perfectly preserved within its conditioned slices.

The conditional self-similarity of random states is robust and preserved under the global
depolarizing noise. The noisy bit-string probability follows the affine map, 
$\tilde{p}(y,z) = (1-\lambda) p(y,z) + {\lambda}/{N}$ and the marginal probability of the
system $B$ is written a $\tilde{p}(b) = \sum_y \tilde{p}(y,b) = (1-\lambda)p(b) + \lambda{M}/{N}$.
Thus, the conditional probability is given by
\begin{subequations}
\begin{align}
\tilde{p}_{A|b} &= \frac{q(y,b)}{\tilde{q}(b)} 
                = \frac{(1-\lambda) p(y,b) + \lambda/N}{(1-\lambda)p(b) + \lambda M/N}\\
                &\approx (1-\lambda) p_{A|b} + \frac{\lambda}{M} \,,
\label{Eq:Affine_sale}
\end{align}
\end{subequations}
where the typicality approximation $p(b) \approx {\mathbb E}[p(b)] = M/N$ is used and valid in 
the limit of $M\gg N-M$. Eq.~(\ref{Eq:Affine_sale}) is identical to 
Eq.~(\ref{Eq:depolarized_state_affine}) by replacing $p$ and $N$ with $p_{A|b}$ and $M$, 
respectively. 

To validate the conditional scale invariance of random circuits, we analyze Google’s Sycamore 
sampling data for $n=12$ qubits~\cite{Martinis2022}. Fig.\ref{Fig:conditional} displays 
the empirical distributions for the conditional probabilities $p(y|b=0)$ (conditioned on the 
final bit) and $p(y|b=00)$ (conditioned on the final two bits), contrasted with the marginalized 
distribution for $p(y)$ obtained by tracing out the last two qubits. Consistent with our theory, 
the PDFs for $p(y|b=0)$ and $p(y|b=00)$ are statistically indistinguishable, confirming that 
the depolarizing gap $\lambda$ is conserved across different conditioning scales. 
In contrast, the marginalized PDF $p(y)$ deviates significantly from the pure-state statistics. 
The solid lines in Fig. 3 represent the scaled-shifted exponential distribution 
Eq.~(\ref{Eq:scaled_shifted_exponential}), which shows excellent agreement with the conditioned 
experimental data. Minor discrepancies between Fig. .~\ref{Fig:conditional} (a) and (b) are 
attributed to finite sample sizes, while the subtle smearing of the depolarizing gap
originates from secondary noise sources—such as read-out errors beyond depolarizing model,
as discussed before.

\paragraph{Subsystem XEB or Conditional XEB.---} 
The conditional scale invariance of random states provides a scalable framework for validating 
random circuit sampling. The full-system XEB~\cite{Boixo2018,Arute2019} quantifies fidelity by 
correlating experimental samples $\{x_i\}$ with ideal probabilities $p_{\text{ideal}}(x)$ 
through the linear estimator $F_{\text{XEB}} = 2^n \langle p_{\text{ideal}}(x_i) \rangle - 1$. 
Similarly, the subsystem XEB can be defined as $F_{\text{XEB}}^{(A)} 
= 2^{m} \langle p_{\text{ideal}}^{(A)}(y_i) \rangle - 1$ where 
$m$ is the size of the subsystem $A$, $\{y_i\}$ its bit-strings, and 
$p_{\rm ideal}^{(A)}$ the Beta distribution, Eq.~{(\ref{Eq:Sub_Beta_x})}.
Our discovery of conditional self-similarity enables the more rigorous conditional XEB:
\begin{equation}
F_{\text{XEB}}^{(A|b)} = 2^{m} \langle p_{\text{ideal}}(y_i |b) \rangle - 1,
\end{equation}
where the conditional probabilities $p_{\text{ideal}}(y_i | b)$ follow the universal distribution 
Eq.~{(\ref{Eq:Cond_Beta_p})}. The primary advantage of this approach is computational scalability: 
the conditioned subsystem with $m\ll n$ can be simulated with less computational 
costs than the full system.

In summary, we analytically derived the distribution of subsystem bit-string probabilities for 
both Haar-random pure states and depolarized random states within a unified Dirichlet framework, 
showing that subsystem probabilities follow Beta distributions obtained via Dirichlet aggregation. 
For the full system, the Beta law reduces to the exponential distribution, while for subsystems 
it interpolates toward Gamma or Gaussian limits depending on subsystem size. Depolarizing noise 
acts as an affine transformation on the Dirichlet-distributed probability vector, producing 
shifted and rescaled Beta distributions with a characteristic depolarizing gap.

More fundamentally, we prove that Haar random states exhibit exact conditional scale invariance 
or conditional self-similarity, arising form the neutrality of the Dirichlet distribution.
This property is mathematically exact across all subsystem sizes, revealing a fundamental 
symmetry hidden beneath the apparent randomness of Hilbert space. It shows that Haar randomness 
is not merely global disorder, but possesses a recursive statistical structure: any conditioned 
subsystem mirrors the statistics of the full system at reduced dimension.

The analytic distributions of subsystems and conditional subsystems derived in this work establish 
subsystem and conditional cross-entropy benchmarking as a powerful and scalable framework
for random circuit sampling. By restricting analysis on reduced subsystems, this approach can s
substantially lower the classical computational cost of evaluating ideal probabilities while preserving 
sensitivity to the underlying statistical structure. Moreover, conditional XEB provides 
a stringent test against classical spoofing strategies~\cite{Pan2022}, as it probes not only marginal 
distributions but also the nontrivial conditional dependencies characteristic of 
genuine Haar-random states.
\acknowledgements
\bibliography{PM_Statistics.bib}

\begin{thebibliography}{31}%
\makeatletter
\providecommand \@ifxundefined [1]{%
 \@ifx{#1\undefined}
}%
\providecommand \@ifnum [1]{%
 \ifnum #1\expandafter \@firstoftwo
 \else \expandafter \@secondoftwo
 \fi
}%
\providecommand \@ifx [1]{%
 \ifx #1\expandafter \@firstoftwo
 \else \expandafter \@secondoftwo
 \fi
}%
\providecommand \natexlab [1]{#1}%
\providecommand \enquote  [1]{``#1''}%
\providecommand \bibnamefont  [1]{#1}%
\providecommand \bibfnamefont [1]{#1}%
\providecommand \citenamefont [1]{#1}%
\providecommand \href@noop [0]{\@secondoftwo}%
\providecommand \href [0]{\begingroup \@sanitize@url \@href}%
\providecommand \@href[1]{\@@startlink{#1}\@@href}%
\providecommand \@@href[1]{\endgroup#1\@@endlink}%
\providecommand \@sanitize@url [0]{\catcode `\\12\catcode `\$12\catcode
  `\&12\catcode `\#12\catcode `\^12\catcode `\_12\catcode `\%12\relax}%
\providecommand \@@startlink[1]{}%
\providecommand \@@endlink[0]{}%
\providecommand \url  [0]{\begingroup\@sanitize@url \@url }%
\providecommand \@url [1]{\endgroup\@href {#1}{\urlprefix }}%
\providecommand \urlprefix  [0]{URL }%
\providecommand \Eprint [0]{\href }%
\providecommand \doibase [0]{https://doi.org/}%
\providecommand \selectlanguage [0]{\@gobble}%
\providecommand \bibinfo  [0]{\@secondoftwo}%
\providecommand \bibfield  [0]{\@secondoftwo}%
\providecommand \translation [1]{[#1]}%
\providecommand \BibitemOpen [0]{}%
\providecommand \bibitemStop [0]{}%
\providecommand \bibitemNoStop [0]{.\EOS\space}%
\providecommand \EOS [0]{\spacefactor3000\relax}%
\providecommand \BibitemShut  [1]{\csname bibitem#1\endcsname}%
\let\auto@bib@innerbib\@empty
\bibitem [{\citenamefont {Ledoux}(2001)}]{Ledoux2001}%
  \BibitemOpen
  \bibfield  {author} {\bibinfo {author} {\bibfnamefont {M.}~\bibnamefont
  {Ledoux}},\ }\href@noop {} {\emph {\bibinfo {title} {The Concentration of
  Measure Phenomenon}}},\ \bibinfo {series} {Mathematical Surveys and
  Monographs}, Vol.~\bibinfo {volume} {89}\ (\bibinfo  {publisher} {American
  Mathematical Society},\ \bibinfo {address} {Providence, R.I.},\ \bibinfo
  {year} {2001})\BibitemShut {NoStop}%
\bibitem [{\citenamefont {Hayden}\ \emph {et~al.}(2004)\citenamefont {Hayden},
  \citenamefont {Leung}, \citenamefont {Shor},\ and\ \citenamefont
  {Winter}}]{Hayden2004}%
  \BibitemOpen
  \bibfield  {author} {\bibinfo {author} {\bibfnamefont {P.}~\bibnamefont
  {Hayden}}, \bibinfo {author} {\bibfnamefont {D.}~\bibnamefont {Leung}},
  \bibinfo {author} {\bibfnamefont {P.~W.}\ \bibnamefont {Shor}},\ and\
  \bibinfo {author} {\bibfnamefont {A.}~\bibnamefont {Winter}},\ }\bibfield
  {title} {\bibinfo {title} {Randomizing quantum states: Constructions and
  applications},\ }\href {https://doi.org/10.1007/s00220-004-1087-6} {\bibfield
   {journal} {\bibinfo  {journal} {Communications in Mathematical Physics}\
  }\textbf {\bibinfo {volume} {250}},\ \bibinfo {pages} {371} (\bibinfo {year}
  {2004})}\BibitemShut {NoStop}%
\bibitem [{\citenamefont {Page}(1993{\natexlab{a}})}]{Page1993a}%
  \BibitemOpen
  \bibfield  {author} {\bibinfo {author} {\bibfnamefont {D.~N.}\ \bibnamefont
  {Page}},\ }\bibfield  {title} {\bibinfo {title} {Average entropy of a
  subsystem},\ }\href {https://doi.org/10.1103/PhysRevLett.71.1291} {\bibfield
  {journal} {\bibinfo  {journal} {Phys. Rev. Lett.}\ }\textbf {\bibinfo
  {volume} {71}},\ \bibinfo {pages} {1291} (\bibinfo {year}
  {1993}{\natexlab{a}})}\BibitemShut {NoStop}%
\bibitem [{\citenamefont {Popescu}\ \emph {et~al.}(2006)\citenamefont
  {Popescu}, \citenamefont {Short},\ and\ \citenamefont
  {Winter}}]{Popescu2006}%
  \BibitemOpen
  \bibfield  {author} {\bibinfo {author} {\bibfnamefont {S.}~\bibnamefont
  {Popescu}}, \bibinfo {author} {\bibfnamefont {A.~J.}\ \bibnamefont {Short}},\
  and\ \bibinfo {author} {\bibfnamefont {A.}~\bibnamefont {Winter}},\
  }\bibfield  {title} {\bibinfo {title} {Entanglement and the foundations of
  statistical mechanics},\ }\href {https://doi.org/10.1038/nphys444} {\bibfield
   {journal} {\bibinfo  {journal} {Nature Physics}\ }\textbf {\bibinfo {volume}
  {2}},\ \bibinfo {pages} {754} (\bibinfo {year} {2006})}\BibitemShut {NoStop}%
\bibitem [{\citenamefont {Reimann}(2008)}]{Reimann2008}%
  \BibitemOpen
  \bibfield  {author} {\bibinfo {author} {\bibfnamefont {P.}~\bibnamefont
  {Reimann}},\ }\bibfield  {title} {\bibinfo {title} {Typicality of pure states
  randomly sampled according to the gaussian adjusted projected measure},\
  }\href {https://doi.org/10.1007/s10955-008-9576-1} {\bibfield  {journal}
  {\bibinfo  {journal} {Journal of Statistical Physics}\ }\textbf {\bibinfo
  {volume} {132}},\ \bibinfo {pages} {921} (\bibinfo {year}
  {2008})}\BibitemShut {NoStop}%
\bibitem [{\citenamefont {Ku\'s}\ \emph {et~al.}(1988)\citenamefont {Ku\'s},
  \citenamefont {Mostowski},\ and\ \citenamefont {Haake}}]{Kus1988}%
  \BibitemOpen
  \bibfield  {author} {\bibinfo {author} {\bibfnamefont {M.}~\bibnamefont
  {Ku\'s}}, \bibinfo {author} {\bibfnamefont {J.}~\bibnamefont {Mostowski}},\
  and\ \bibinfo {author} {\bibfnamefont {F.}~\bibnamefont {Haake}},\ }\bibfield
   {title} {\bibinfo {title} {Universality of eigenvector statistics of kicked
  tops of different symmetries},\ }\href
  {https://doi.org/10.1088/0305-4470/21/22/006} {\bibfield  {journal} {\bibinfo
   {journal} {Journal of Physics A: Mathematical and General}\ }\textbf
  {\bibinfo {volume} {21}},\ \bibinfo {pages} {L1073} (\bibinfo {year}
  {1988})}\BibitemShut {NoStop}%
\bibitem [{\citenamefont {Haake}\ \emph {et~al.}(2010)\citenamefont {Haake},
  \citenamefont {Gnutzmann},\ and\ \citenamefont {Ku\'s}}]{Haake2010}%
  \BibitemOpen
  \bibfield  {author} {\bibinfo {author} {\bibfnamefont {F.}~\bibnamefont
  {Haake}}, \bibinfo {author} {\bibfnamefont {S.}~\bibnamefont {Gnutzmann}},\
  and\ \bibinfo {author} {\bibfnamefont {M.}~\bibnamefont {Ku\'s}},\ }\href
  {https://doi.org/10.1007/978-3-642-05428-0} {\emph {\bibinfo {title} {Quantum
  Signatures of Chaos}}}\ (\bibinfo  {publisher} {Springer-Verlag Berlin
  Heidelberg},\ \bibinfo {year} {2010})\BibitemShut {NoStop}%
\bibitem [{\citenamefont {Brody}\ \emph {et~al.}(1981)\citenamefont {Brody},
  \citenamefont {Flores}, \citenamefont {French}, \citenamefont {Mello},
  \citenamefont {Pandey},\ and\ \citenamefont {Wong}}]{Broady1981}%
  \BibitemOpen
  \bibfield  {author} {\bibinfo {author} {\bibfnamefont {T.~A.}\ \bibnamefont
  {Brody}}, \bibinfo {author} {\bibfnamefont {J.}~\bibnamefont {Flores}},
  \bibinfo {author} {\bibfnamefont {J.~B.}\ \bibnamefont {French}}, \bibinfo
  {author} {\bibfnamefont {P.~A.}\ \bibnamefont {Mello}}, \bibinfo {author}
  {\bibfnamefont {A.}~\bibnamefont {Pandey}},\ and\ \bibinfo {author}
  {\bibfnamefont {S.~S.~M.}\ \bibnamefont {Wong}},\ }\bibfield  {title}
  {\bibinfo {title} {Random-matrix physics: spectrum and strength
  fluctuations},\ }\href {https://doi.org/10.1103/RevModPhys.53.385} {\bibfield
   {journal} {\bibinfo  {journal} {Rev. Mod. Phys.}\ }\textbf {\bibinfo
  {volume} {53}},\ \bibinfo {pages} {385} (\bibinfo {year} {1981})}\BibitemShut
  {NoStop}%
\bibitem [{\citenamefont {Mehta}(2004)}]{Mehta2004}%
  \BibitemOpen
  \bibfield  {author} {\bibinfo {author} {\bibfnamefont {M.~L.}\ \bibnamefont
  {Mehta}},\ }\href
  {https://doi.org/https://doi.org/10.1016/S0079-8169(13)62915-3} {\emph
  {\bibinfo {title} {Random Matrices}}},\ \bibinfo {series} {Pure and Applied
  Mathematics}, Vol.\ \bibinfo {volume} {142}\ (\bibinfo  {publisher}
  {Elsevier},\ \bibinfo {year} {2004})\BibitemShut {NoStop}%
\bibitem [{\citenamefont {Page}(1993{\natexlab{b}})}]{Page1993b}%
  \BibitemOpen
  \bibfield  {author} {\bibinfo {author} {\bibfnamefont {D.~N.}\ \bibnamefont
  {Page}},\ }\bibfield  {title} {\bibinfo {title} {Information in black hole
  radiation},\ }\href {https://doi.org/10.1103/PhysRevLett.71.3743} {\bibfield
  {journal} {\bibinfo  {journal} {Phys. Rev. Lett.}\ }\textbf {\bibinfo
  {volume} {71}},\ \bibinfo {pages} {3743} (\bibinfo {year}
  {1993}{\natexlab{b}})}\BibitemShut {NoStop}%
\bibitem [{\citenamefont {Hayden}\ and\ \citenamefont
  {Preskill}(2007)}]{Hayden2007}%
  \BibitemOpen
  \bibfield  {author} {\bibinfo {author} {\bibfnamefont {P.}~\bibnamefont
  {Hayden}}\ and\ \bibinfo {author} {\bibfnamefont {J.}~\bibnamefont
  {Preskill}},\ }\bibfield  {title} {\bibinfo {title} {Black holes as mirrors:
  quantum information in random subsystems},\ }\href
  {https://doi.org/10.1088/1126-6708/2007/09/120} {\bibfield  {journal}
  {\bibinfo  {journal} {Journal of High Energy Physics}\ }\textbf {\bibinfo
  {volume} {2007}},\ \bibinfo {pages} {120} (\bibinfo {year}
  {2007})}\BibitemShut {NoStop}%
\bibitem [{\citenamefont {Boixo}\ \emph {et~al.}(2018)\citenamefont {Boixo},
  \citenamefont {Isakov}, \citenamefont {Smelyanskiy}, \citenamefont {Babbush},
  \citenamefont {Ding}, \citenamefont {Jiang}, \citenamefont {Bremner},
  \citenamefont {Martinis},\ and\ \citenamefont {Neven}}]{Boixo2018}%
  \BibitemOpen
  \bibfield  {author} {\bibinfo {author} {\bibfnamefont {S.}~\bibnamefont
  {Boixo}}, \bibinfo {author} {\bibfnamefont {S.~V.}\ \bibnamefont {Isakov}},
  \bibinfo {author} {\bibfnamefont {V.~N.}\ \bibnamefont {Smelyanskiy}},
  \bibinfo {author} {\bibfnamefont {R.}~\bibnamefont {Babbush}}, \bibinfo
  {author} {\bibfnamefont {N.}~\bibnamefont {Ding}}, \bibinfo {author}
  {\bibfnamefont {Z.}~\bibnamefont {Jiang}}, \bibinfo {author} {\bibfnamefont
  {M.~J.}\ \bibnamefont {Bremner}}, \bibinfo {author} {\bibfnamefont {J.~M.}\
  \bibnamefont {Martinis}},\ and\ \bibinfo {author} {\bibfnamefont
  {H.}~\bibnamefont {Neven}},\ }\bibfield  {title} {\bibinfo {title}
  {Characterizing quantum supremacy in near-term devices},\ }\href
  {https://doi.org/10.1038/s41567-018-0124-x} {\bibfield  {journal} {\bibinfo
  {journal} {Nature Physics}\ }\textbf {\bibinfo {volume} {14}},\ \bibinfo
  {pages} {595} (\bibinfo {year} {2018})}\BibitemShut {NoStop}%
\bibitem [{\citenamefont {Arute}\ \emph {et~al.}(2019)\citenamefont {Arute}
  \emph {et~al.}}]{Arute2019}%
  \BibitemOpen
  \bibfield  {author} {\bibinfo {author} {\bibfnamefont {F.}~\bibnamefont
  {Arute}} \emph {et~al.},\ }\bibfield  {title} {\bibinfo {title} {Quantum
  supremacy using a programmable superconducting processor},\ }\href
  {https://doi.org/10.1038/s41586-019-1666-5} {\bibfield  {journal} {\bibinfo
  {journal} {Nature}\ }\textbf {\bibinfo {volume} {574}},\ \bibinfo {pages}
  {505} (\bibinfo {year} {2019})}\BibitemShut {NoStop}%
\bibitem [{\citenamefont {Bouland}\ \emph {et~al.}(2019)\citenamefont
  {Bouland}, \citenamefont {Fefferman}, \citenamefont {Nirkhe},\ and\
  \citenamefont {Vazirani}}]{Bouland2019}%
  \BibitemOpen
  \bibfield  {author} {\bibinfo {author} {\bibfnamefont {A.}~\bibnamefont
  {Bouland}}, \bibinfo {author} {\bibfnamefont {B.}~\bibnamefont {Fefferman}},
  \bibinfo {author} {\bibfnamefont {C.}~\bibnamefont {Nirkhe}},\ and\ \bibinfo
  {author} {\bibfnamefont {U.}~\bibnamefont {Vazirani}},\ }\bibfield  {title}
  {\bibinfo {title} {On the complexity and verification of quantum random
  circuit sampling},\ }\href {https://doi.org/10.1038/s41567-018-0318-2}
  {\bibfield  {journal} {\bibinfo  {journal} {Nature Physics}\ }\textbf
  {\bibinfo {volume} {15}},\ \bibinfo {pages} {159} (\bibinfo {year}
  {2019})}\BibitemShut {NoStop}%
\bibitem [{\citenamefont {Wu}\ \emph {et~al.}(2021)\citenamefont {Wu} \emph
  {et~al.}}]{Wu2021}%
  \BibitemOpen
  \bibfield  {author} {\bibinfo {author} {\bibfnamefont {Y.}~\bibnamefont {Wu}}
  \emph {et~al.},\ }\bibfield  {title} {\bibinfo {title} {Strong quantum
  computational advantage using a superconducting quantum processor},\ }\href
  {https://doi.org/10.1103/PhysRevLett.127.180501} {\bibfield  {journal}
  {\bibinfo  {journal} {Phys. Rev. Lett.}\ }\textbf {\bibinfo {volume} {127}},\
  \bibinfo {pages} {180501} (\bibinfo {year} {2021})}\BibitemShut {NoStop}%
\bibitem [{\citenamefont {Zhu}\ \emph {et~al.}(2022)\citenamefont {Zhu} \emph
  {et~al.}}]{Zhu2022}%
  \BibitemOpen
  \bibfield  {author} {\bibinfo {author} {\bibfnamefont {Q.}~\bibnamefont
  {Zhu}} \emph {et~al.},\ }\bibfield  {title} {\bibinfo {title} {Quantum
  computational advantage via 60-qubit 24-cycle random circuit sampling},\
  }\href {https://doi.org/https://doi.org/10.1016/j.scib.2021.10.017}
  {\bibfield  {journal} {\bibinfo  {journal} {Science Bulletin}\ }\textbf
  {\bibinfo {volume} {67}},\ \bibinfo {pages} {240} (\bibinfo {year}
  {2022})}\BibitemShut {NoStop}%
\bibitem [{\citenamefont {Hangleiter}\ and\ \citenamefont
  {Eisert}(2023)}]{Hangleiter2023}%
  \BibitemOpen
  \bibfield  {author} {\bibinfo {author} {\bibfnamefont {D.}~\bibnamefont
  {Hangleiter}}\ and\ \bibinfo {author} {\bibfnamefont {J.}~\bibnamefont
  {Eisert}},\ }\bibfield  {title} {\bibinfo {title} {Computational advantage of
  quantum random sampling},\ }\href
  {https://doi.org/10.1103/RevModPhys.95.035001} {\bibfield  {journal}
  {\bibinfo  {journal} {Rev. Mod. Phys.}\ }\textbf {\bibinfo {volume} {95}},\
  \bibinfo {pages} {035001} (\bibinfo {year} {2023})}\BibitemShut {NoStop}%
\bibitem [{\citenamefont {Morvan}\ \emph {et~al.}(2024)\citenamefont {Morvan}
  \emph {et~al.}}]{Morvan2024}%
  \BibitemOpen
  \bibfield  {author} {\bibinfo {author} {\bibfnamefont {A.}~\bibnamefont
  {Morvan}} \emph {et~al.},\ }\bibfield  {title} {\bibinfo {title} {Phase
  transitions in random circuit sampling},\ }\href
  {https://doi.org/10.1038/s41586-024-07998-6} {\bibfield  {journal} {\bibinfo
  {journal} {Nature}\ }\textbf {\bibinfo {volume} {634}},\ \bibinfo {pages}
  {328} (\bibinfo {year} {2024})}\BibitemShut {NoStop}%
\bibitem [{\citenamefont {Gao}\ \emph {et~al.}(2025)\citenamefont {Gao} \emph
  {et~al.}}]{Gao2025}%
  \BibitemOpen
  \bibfield  {author} {\bibinfo {author} {\bibfnamefont {D.}~\bibnamefont
  {Gao}} \emph {et~al.},\ }\bibfield  {title} {\bibinfo {title} {Establishing a
  new benchmark in quantum computational advantage with 105-qubit zuchongzhi
  3.0 processor},\ }\href {https://doi.org/10.1103/PhysRevLett.134.090601}
  {\bibfield  {journal} {\bibinfo  {journal} {Phys. Rev. Lett.}\ }\textbf
  {\bibinfo {volume} {134}},\ \bibinfo {pages} {090601} (\bibinfo {year}
  {2025})}\BibitemShut {NoStop}%
\bibitem [{\citenamefont {Liu}\ \emph {et~al.}(2025)\citenamefont {Liu} \emph
  {et~al.}}]{Liu2025}%
  \BibitemOpen
  \bibfield  {author} {\bibinfo {author} {\bibfnamefont {M.}~\bibnamefont
  {Liu}} \emph {et~al.},\ }\bibfield  {title} {\bibinfo {title} {Certified
  randomness using a trapped-ion quantum processor},\ }\href
  {https://doi.org/10.1038/s41586-025-08737-1} {\bibfield  {journal} {\bibinfo
  {journal} {Nature}\ }\textbf {\bibinfo {volume} {640}},\ \bibinfo {pages}
  {343} (\bibinfo {year} {2025})}\BibitemShut {NoStop}%
\bibitem [{\citenamefont {Aaronson}\ and\ \citenamefont
  {Gunn}(2020)}]{Aaronson2020}%
  \BibitemOpen
  \bibfield  {author} {\bibinfo {author} {\bibfnamefont {S.}~\bibnamefont
  {Aaronson}}\ and\ \bibinfo {author} {\bibfnamefont {S.}~\bibnamefont
  {Gunn}},\ }\href {https://arxiv.org/abs/1910.12085} {\bibinfo {title} {On the
  classical hardness of spoofing linear cross-entropy benchmarking}} (\bibinfo
  {year} {2020}),\ \Eprint {https://arxiv.org/abs/1910.12085} {arXiv:1910.12085
  [quant-ph]} \BibitemShut {NoStop}%
\bibitem [{\citenamefont {Balakrishnan}\ and\ \citenamefont
  {Nevzorov}(2004)}]{Balakrishnan2004}%
  \BibitemOpen
  \bibfield  {author} {\bibinfo {author} {\bibfnamefont {N.}~\bibnamefont
  {Balakrishnan}}\ and\ \bibinfo {author} {\bibfnamefont {V.}~\bibnamefont
  {Nevzorov}},\ }\href {https://books.google.com/books?id=JIfk5kBdLGIC} {\emph
  {\bibinfo {title} {A Primer on Statistical Distributions}}}\ (\bibinfo
  {publisher} {Wiley},\ \bibinfo {year} {2004})\BibitemShut {NoStop}%
\bibitem [{\citenamefont {Zyczkowski}\ and\ \citenamefont
  {Sommers}(2001)}]{Zyczkowski2001}%
  \BibitemOpen
  \bibfield  {author} {\bibinfo {author} {\bibfnamefont {K.}~\bibnamefont
  {Zyczkowski}}\ and\ \bibinfo {author} {\bibfnamefont {H.-J.}\ \bibnamefont
  {Sommers}},\ }\bibfield  {title} {\bibinfo {title} {Induced measures in the
  space of mixed quantum states},\ }\href
  {https://doi.org/10.1088/0305-4470/34/35/335} {\bibfield  {journal} {\bibinfo
   {journal} {Journal of Physics A: Mathematical and General}\ }\textbf
  {\bibinfo {volume} {34}},\ \bibinfo {pages} {7111} (\bibinfo {year}
  {2001})}\BibitemShut {NoStop}%
\bibitem [{\citenamefont {Bengtsson}\ and\ \citenamefont
  {Zyczkowski}(2007)}]{Bengtsson2007}%
  \BibitemOpen
  \bibfield  {author} {\bibinfo {author} {\bibfnamefont {I.}~\bibnamefont
  {Bengtsson}}\ and\ \bibinfo {author} {\bibfnamefont {K.}~\bibnamefont
  {Zyczkowski}},\ }\href {https://books.google.com/books?id=aA4vXMbuOTUC}
  {\emph {\bibinfo {title} {Geometry of Quantum States: An Introduction to
  Quantum Entanglement}}}\ (\bibinfo  {publisher} {Cambridge University
  Press},\ \bibinfo {year} {2007})\BibitemShut {NoStop}%
\bibitem [{\citenamefont {Kotz}\ \emph {et~al.}(2019)\citenamefont {Kotz},
  \citenamefont {Balakrishnan},\ and\ \citenamefont {Johnson}}]{Kotz2019}%
  \BibitemOpen
  \bibfield  {author} {\bibinfo {author} {\bibfnamefont {S.}~\bibnamefont
  {Kotz}}, \bibinfo {author} {\bibfnamefont {N.}~\bibnamefont {Balakrishnan}},\
  and\ \bibinfo {author} {\bibfnamefont {N.~L.}\ \bibnamefont {Johnson}},\
  }\href@noop {} {\emph {\bibinfo {title} {Continuous multivariate
  distributions, Volume 1: Models and applications}}},\ Vol.~\bibinfo {volume}
  {1}\ (\bibinfo  {publisher} {John wiley \& sons},\ \bibinfo {year}
  {2019})\BibitemShut {NoStop}%
\bibitem [{\citenamefont {Martinis~\textit{et~al.}}(2022)}]{Martinis2022}%
  \BibitemOpen
  \bibfield  {author} {\bibinfo {author} {\bibfnamefont {J.~M.}\ \bibnamefont
  {Martinis~\textit{et~al.}}},\ }\href
  {https://doi.org/https://doi.org/10.5061/dryad.k6t1rj8} {\bibinfo {title}
  {Quantum supremacy using a programmable superconducting processor, {D}ryad,
  {D}ataset}},\ \bibinfo {howpublished}
  {\url{https://doi.org/10.5061/dryad.k6t1rj8}} (\bibinfo {year}
  {2022})\BibitemShut {NoStop}%
\bibitem [{\citenamefont {Oh}\ and\ \citenamefont {Kais}(2022)}]{Oh2022b}%
  \BibitemOpen
  \bibfield  {author} {\bibinfo {author} {\bibfnamefont {S.}~\bibnamefont
  {Oh}}\ and\ \bibinfo {author} {\bibfnamefont {S.}~\bibnamefont {Kais}},\
  }\bibfield  {title} {\bibinfo {title} {Statistical properties of bit strings
  sampled from sycamore random quantum circuits},\ }\href
  {https://doi.org/10.1021/acs.jpclett.2c02045} {\bibfield  {journal} {\bibinfo
   {journal} {The Journal of Physical Chemistry Letters}\ }\textbf {\bibinfo
  {volume} {13}},\ \bibinfo {pages} {7469} (\bibinfo {year}
  {2022})}\BibitemShut {NoStop}%
\bibitem [{\citenamefont {Oh}()}]{Oh2026}%
  \BibitemOpen
  \bibfield  {author} {\bibinfo {author} {\bibfnamefont {S.}~\bibnamefont
  {Oh}},\ }\href@noop {} {\bibinfo {title} {Analysis of readout errors in
  random circuit sampling (in preparation)}}\BibitemShut {NoStop}%
\bibitem [{\citenamefont {Lukacs}(1955)}]{Lukacs1955}%
  \BibitemOpen
  \bibfield  {author} {\bibinfo {author} {\bibfnamefont {E.}~\bibnamefont
  {Lukacs}},\ }\bibfield  {title} {\bibinfo {title} {A characterization of the
  gamma distribution},\ }\href@noop {} {\bibfield  {journal} {\bibinfo
  {journal} {The Annals of Mathematical Statistics}\ }\textbf {\bibinfo
  {volume} {26}},\ \bibinfo {pages} {319} (\bibinfo {year} {1955})}\BibitemShut
  {NoStop}%
\bibitem [{\citenamefont {Connor}\ and\ \citenamefont
  {Mosimann}(1969)}]{Connor1969}%
  \BibitemOpen
  \bibfield  {author} {\bibinfo {author} {\bibfnamefont {R.~J.}\ \bibnamefont
  {Connor}}\ and\ \bibinfo {author} {\bibfnamefont {J.~E.}\ \bibnamefont
  {Mosimann}},\ }\bibfield  {title} {\bibinfo {title} {Concepts of independence
  for proportions with a generalization of the dirichlet distribution},\
  }\href@noop {} {\bibfield  {journal} {\bibinfo  {journal} {Journal of the
  American Statistical Association}\ }\textbf {\bibinfo {volume} {64}},\
  \bibinfo {pages} {194} (\bibinfo {year} {1969})}\BibitemShut {NoStop}%
\bibitem [{\citenamefont {Pan}\ and\ \citenamefont {Zhang}(2022)}]{Pan2022}%
  \BibitemOpen
  \bibfield  {author} {\bibinfo {author} {\bibfnamefont {F.}~\bibnamefont
  {Pan}}\ and\ \bibinfo {author} {\bibfnamefont {P.}~\bibnamefont {Zhang}},\
  }\bibfield  {title} {\bibinfo {title} {Simulation of quantum circuits using
  the big-batch tensor network method},\ }\href
  {https://doi.org/10.1103/PhysRevLett.128.030501} {\bibfield  {journal}
  {\bibinfo  {journal} {Phys. Rev. Lett.}\ }\textbf {\bibinfo {volume} {128}},\
  \bibinfo {pages} {030501} (\bibinfo {year} {2022})}\BibitemShut {NoStop}%
\end{thebibliography}%
\end{document}